\documentstyle[11pt,aaspp4]{article}

\received{9-Oct-1996}
\accepted{21-Nov-1996}

\newcommand{\CI}{\mbox{[CI]}}
\newcommand{\threePtwoone}{\mbox{$^{3}P_{2}\!\rightarrow\!\!^{3}P_{1}$}}
\newcommand{\threePonezero}{\mbox{$^{3}P_{1}\!\rightarrow\!\!^{3}P_{0}$}}
\newcommand{\cmcube}{\mbox{\rm cm$^{-3}$}}
\newcommand{\cmsqu}{\mbox{\rm cm$^{-2}$}}
\newcommand{\Jtwoone}{\mbox{$J\!=\!2\!\rightarrow\!1$}}
\newcommand{\Jonezero}{\mbox{$J\!=\!1\!\rightarrow\!0$}}
\newcommand{\Htwo}{\mbox{\rm H$_2$}}
\newcommand{\CeighteenO}{\mbox{\rm C$^{18}$O}}
\newcommand{\kms}{\mbox{\rm km s$^{-1}$}}
\newcommand{\Kkms}{\mbox{[\rm K\,km\,s$^{-1}$]}}

\lefthead{Stutzki et al.}
\righthead{Atomic Carbon in M\,82}

\begin{document}

    \title{
    Atomic Carbon in M\,82:\\
    Physical conditions derived
    from simultaneous observations of the [CI] fine structure
    submillimeter wave transitions}

    \author{
     J. Stutzki, 
     U.U. Graf, 
     S. Haas, 
     C.E. Honingh, 
     D. Hottgenroth, 
     K. Jacobs, 
     R. Schieder, 
     R. Simon, 
     J. Staguhn and  
     G. Winnewisser}
\affil{    
     I. Physikalisches Institut der Universit\"at zu K\"oln,
     Z\"ulpicher Stra{\ss}e 77, D-50937 K\"oln, Germany
		 }
\and 
     \author{R.N. Martin, W.L. Peters, and
     J.P. McMullin}
\affil{ Steward Observatory, The University of Arizona, Tucson, AZ 85721, 
USA}

\begin{abstract}
We report the first extragalactic detection of the neutral carbon [CI] 
\threePtwoone\ 
fine structure line
at 809 GHz. The line was 
observed towards M\,82
simultaneously with the \threePonezero\ 
line at 492 
GHz, providing a precise measurement of 
the $ \Jtwoone / \Jonezero $ 
integrated line
ratio of 0.96 (on a
\Kkms--scale).
This ratio constrains
the [CI] emitting gas to have a temperature of at least 50 K and a density 
of at least $10^4\, \cmcube$. Already at this minimum temperature and 
density, the beam averaged 
CI-column density is large, $2.1\, 10^{18}\,\cmsqu$,
confirming the high CI/CO abundance ratio of $\approx 0.5$ estimated earlier
from the 492 GHz line alone.
We argue that the [CI] emission from M\,82 most likely arises in clouds
of linear size around a few pc with a density of
about $10^4\,\cmcube$ or slightly higher and temperatures of 50\,K up to
about 100\,K.

\end{abstract}


\section{Introduction}
The fine structure lines of neutral atomic carbon, CI, \threePtwoone\ at
809.3435 GHz and \threePonezero\ at 492.1607 GHz (Cooksy et al.\ 
\cite{cooksy:etal}; Yamamoto \& Saito \cite{yama:saito}) provide an important
diagnostic tool for the physical and chemical conditions of the dense
interstellar medium and contribute significantly to
the energy balance of the gas. After the first detections
(\threePonezero: Phillips et al.\ \cite{phill:etal}; \threePtwoone: 
Jaffe et al.\ \cite{jaffe:etal}) the rapid advance in
receiver sensitivity, driven by the technological progress of
superconducting mixers now reaching sensitivities of only a few times
the quantum limit even at submillimeter wavelengths, has resulted
in the 
\Jonezero\ 
line now being well studied 
in many galactic
sources.
The three detections of the \CI\ \threePonezero\ line in external 
galaxies (IC\,342: B\"uttgenbach et al.\ \cite{buett:etal}; M\,82: Schilke et
al.\ \cite{schilke:etal}, White et al.\ \cite{white:etal}; 
NGC\,253: Harrison et al.\ \cite{harrison:etal})
show that CI exhibits rather different properties in these starburst galaxies 
from the ones in the Milky Way as measured by COBE (Wright et 
al.\ \cite{wright:etal}; Bennet et al.\ \cite{benn:etal}).
Both in M\,82 and in NGC\,253 the CI/CO abundance ratio is rather high, 0.4
to 0.5, whereas it is only about 0.1 even in massive star forming regions 
in the Milky Way. 
Observations of the
upper, 
\threePtwoone, 
transition are still rather rare as only few
receiver systems are operational at these wavelengths. 

The \CI\ fine structure line intensity ratio is of particular astrophysical
importance.
In the optically thin regime the integrated line
intensity is proportional to the upper state column density.
The line intensitiy ratio thus directly
measures the \Jtwoone\ excitation temperature,
$\frac{g_2}{g_1}\exp (-h\nu_{21}\!/\!kT_{\rm ex})\,=\,N_2/\!N_1$.
Measuring the integrated line intensity on
a \Kkms--scale,
$R\,=\,\int\!T_{\rm mb,2\rightarrow1}\,dv 
/\!\int\!T_{\rm mb,1\rightarrow0}\,dv$,
and using the values for the A-coefficients given by 
Nussbaumer \& Rusca (\cite{nussb:rusca}),
we obtain
$T_{\rm ex}=38.8 \mbox{ K} /\!\ln \left[2.11 / R\right]$. More
generally,
an escape probability radiative transfer model,  covering also the
case of higher optical depth, links the
observed line ratio with the physical parameters of the source.

In this {\it Letter} we report the first extragalactic detection of the 
\CI\ \threePtwoone\ line, which was observed from the starburst galaxy
M\,82
simultaneouly with the \threePonezero\ line, thus providing a very good 
relative calibration of the line ratio and hence allowing an accurate 
estimate of the excitation conditions in the \CI--emitting gas.

\section{Observations}
The two neutral atomic carbon fine
structure lines 
were observed simultaneously 
using 
a new dual channel receiver 
built at the Universit\"at zu K\"oln (Honingh et al.\ \cite{hon:etal}).
Both channels have a tunerless SIS
waveguide mixer furnished with  SIS junctions designed and fabricated
in the Cologne microstructure laboratory.
One operates from 780 to 820 GHz
in DSB mode with $T_{\rm rec}=1100\,\mbox{ K}$;
the other, from 460-490 GHz, is SSB  
tuned with a
Martin-Puplett interferometer.
Due to technical problems with a cooled IF-amplifier,
the low frequency channel had a $T_{\rm rec}{\rm (SSB)}$ of only
$700\,\mbox{ K}$,
rather poor in comparison with its standard performance
($T_{\rm rec}{\rm (SSB)}=160\,\mbox{ K}$),
but still better than the high frequency
channel and thus not limiting the sensitivity for the simultaneous 
measurement. 
These observations, performed on March 3, 1996,
are the first heterodyne measurements at
frequencies above 460 GHz with
the new 10\,m Heinrich Hertz Telescope 
(Baars et al.\ \cite{baars:etal}) on
Mt. Graham, Arizona, USA.

The receiver package includes
a dual channel acousto-optical spectrometer
backend with a 1.1 GHz bandwidth each and 1\,MHz resolution 
(Schieder et al.\ \cite{schieder:etal}).
Calibration is done via an internal hot load/cold load/sky-flip
mirror, 
following the procedure outlined in 
Harris (\cite{harris:phd}) 
and including
corrections for atmospheric sideband imbalance derived from standard 
atmospheric
models (Grossman \cite{grossman:at}). The 
zenith transmission was 0.41 at 809 GHz and
0.54 
at 492 GHz for an 8 hour period during which the M\,82 data were taken.
Note that 
the ratio of transmissions predicted
by the standard atmospheric model is 0.96.
The difference is likely caused by a
vertical structure of the atmosphere
different from the model assumptions. 

Line intensities are given in units of 
$T_{\rm mb}=T_{\rm A}/(\eta_{\rm mb}\, {\rm e}^{-\tau})$.
Main beam efficiencies were determined according to 
$\eta_{\rm mb}=T_{\rm A,P}/(\eta_{\rm c}\, T_{\rm P}\, {\rm e}^{-\tau})$, where
$T_{\rm A,P}$ is the antenna temperature measured on a planet,
$T_{\rm P}$ the
Rayleigh-Jeans 
temperature of the planet disk, and 
$\eta_{\rm c}\!=\!1\!-\!\exp\, \left[-\ln\!2 \left( d_{\rm P}/
\theta_{\rm FWHM}\right)^2\right]$
the geometric coupling 
factor between the planet (disk diameter $d_{\rm P}$) and the Gaussian beam.
$\eta_{\rm mb}=0.16$ was derived for the 9" FWHM 809 GHz beam, and
$\eta_{\rm mb}=0.29$ for the 15" FWHM 492 GHz beam
from measurements of Venus (17" diameter),
assuming its
physical temperature to be 280\,K 
(Altenhoff \cite{altenhoff:cal}).
Subsequent realignement of the secondary mirror improved these 
efficiencies, but is irrelevant for the data presented here.
Due to the simultaneous observation of both lines, the line
intensity ratio is measured very precisely. Combining the above 
relations, 
the line ratio is effectively calibrated
against the observed 
intensity ratio of the planet and is given by
\begin{eqnarray*}
\frac{T_{\rm mb,1}}{T_{\rm mb,2}} &=&
\frac{T_{\rm A,1}}{T_{\rm A,2}}\,
\left\{ \frac{T_{\rm A,P,2}}{T_{\rm A,P,1}}
\frac{\eta_{\rm c,1}\, T_{\rm P,1}}{\eta_{\rm c,2}\, T_{\rm P,2}} \right\}
\,\exp\left(-(\Delta \tau_{\rm L} - \Delta \tau_{\rm P}) \right),
\end{eqnarray*}
where $\Delta\!\tau=\tau_1\,-\,\tau_2$ is the difference of the atmospheric 
opacity at both line frequencies. Indices L and P distinguish the {\it line}
and {\it planetary} measurements.
As the atmospheric transmission at the \CI\ fine
structure transition frequencies 
is of comparable magnitude, and as long as the planetary calibration
measurement and the source observations are done at roughly the same
transmission, the 
$\exp\left[-(\Delta\!\tau_{\rm L}\,-\,\Delta\!\tau_{\rm P}) \right]$
--factor is always close to unity and depends only 
weakly on the actual transmission. The error in the line ratio is
thus dominated by the
uncertainty in the coupling factor, i.e. the uncertainty  
of the beam sizes at the two frequencies. We did not have a smaller 
planet than Venus available for calibration at the period of the 
observations, and were thus not able to measure the beam sizes directly with 
sufficient precision. Instead, we use the nominal sizes quoted above, 
corresponding to the diffraction limit of a Cassegrain optics with
14 dB edge taper (Goldsmith \cite{gold:qot}). 

The observations were done using the secondary chopping mirror with a
2 arcmin throw at 2 Hz in azimuth and
alternatingly pointing the telescope with the source in one and the
other chopper beam.
As the observations were done in one of the first nights after the
receiver was installed, we did not yet have a good
pointing model available. 
We nevertheless found emission from the SW lobe of M\,82, clearly
identifiable via its line shape and velocity. We mapped several
positions on a 5" grid around the position of strongest emission.
Unfortunately, we were not able to extend our observations to the
center and the NE lobe of M\,82 due to technical problems 
with the telescope drive software after about two 
hours of observing. Pointing
observations over the subsequent days confirmed the M\,82 observations
to be pointed on the SW lobe of the galaxy.

\section{Results and Discussion}
\begin{figure}
\epsscale{0.5}
\plotone{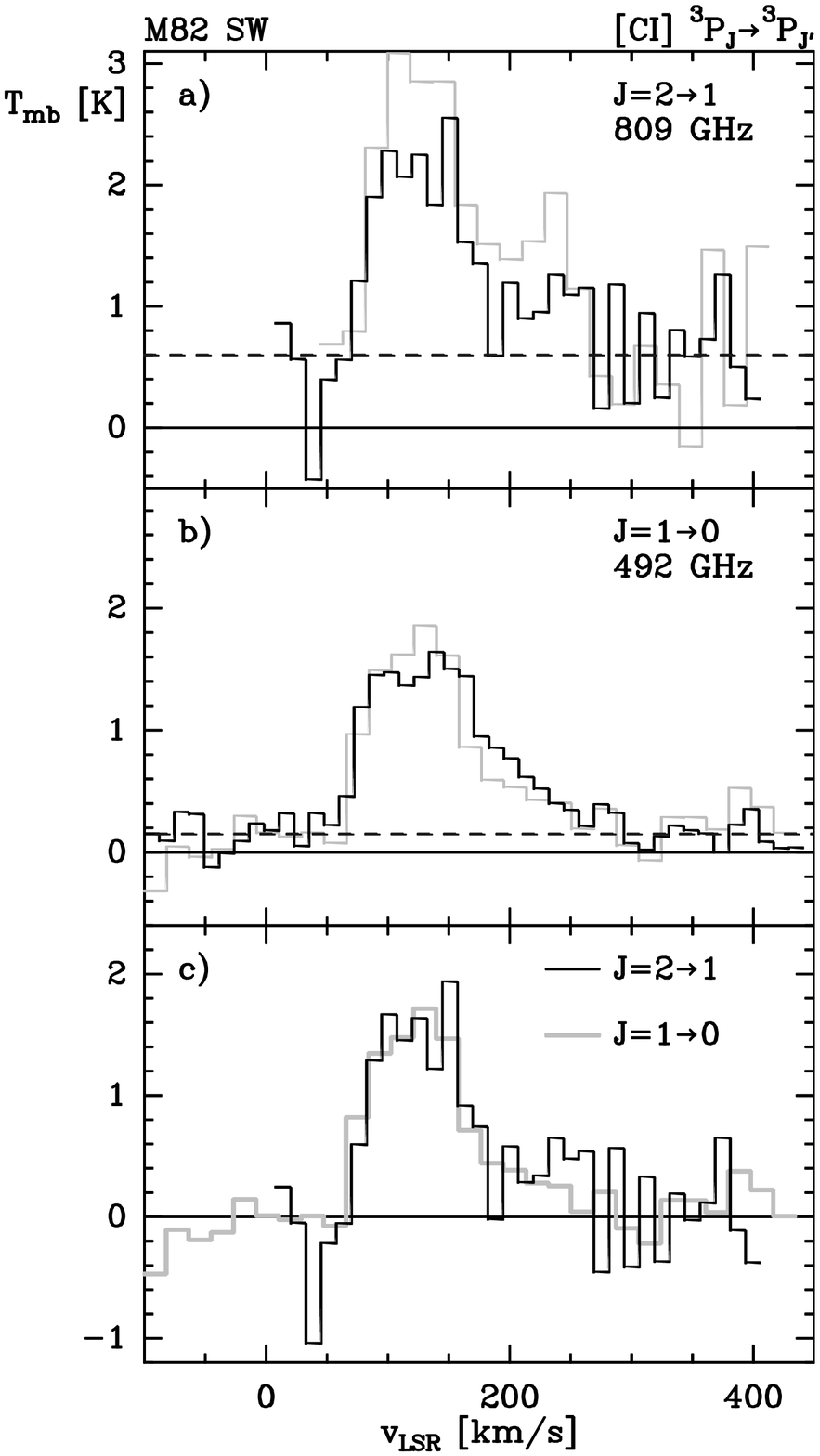}
\caption{
\CI\ spectra of M\,82: a) {\it black}: \CI\ \threePtwoone\ emission
averaged over an
effectively 14" diameter area. {\it Grey}: 
spectrum at the peak of the map. The continuum
offset ({\it dashed line}) 
corresponds to the known dust emission. b) same, 
but for the \CI\ \threePonezero\ line. c) overlay of the continuum
subtracted \CI\ \threePtwoone\ map average and \threePonezero\
peak spectrum.
\label{fig1}
}
\end{figure}

\placefigure{fig1}

Figure 1\,a,b) shows the positionally averaged spectra obtained by summing
33 scans of 2 min duration each, pointed at 8 
positions within a 7" radius from the position of maximum signal. It
thus represents the emission in a beam of effectively
14" FWHM for the 809 GHz and 20" for the 492 GHz line.
Both \CI\ lines are clearly detected with their line shape 
consistent with the earlier observations of the \CI\ 492 GHz emission
from the SW lobe by Schilke et al.\ (\cite{schilke:etal}).
The intensity of the 492 GHz line agrees with that
reported in the same beam size by those authors.
No baseline correction has been
applied to the spectra displayed in panel a) and b). The observed
continuum offsets (0.6 K (DSB) at 809 GHz, i.e. 23 Jy in the effective 
14" beam; 0.15 K (SSB) at 492 GHz, i.e. 8 Jy in the effective 
20" beam) are within the errors
consistent with the broadband dust continuum fluxes 
(450 $\mu$m, Smith et al.\ \cite{smith:etal}; 800 $\mu$m, Hughes, Gear
\& Robson
\cite{hughes:etal};
1300 $\mu$m, Kr\"ugel et al.\ \cite{kruegel:etal}).

At 809 GHz the intensity at the map center 
is significantly
higher than the map average (Fig.\ 1a).
The 
\Jtwoone\ 
line emission is thus more compact than the
map extent. 
The 492 GHz line does not show as
big an effect (Fig.\ 1b),
consistent with it being observed with
the larger beam of 15" FWHM anyway. For the analysis of the line ratio we 
thus compare the 809 GHz map average with the 492 GHz center spectrum 
(Fig.\ 1c). 
The integrated intensity is 152 K \kms\ for the \Jtwoone\ and
158 K \kms\ for
the \Jonezero\ line, 
$R=0.96$, which, 
as discussed above, should be accurate to within 10\%.
The \CI\ intensity ratio in M\,82 is thus much higher
than in the Milky Way: the large
scale COBE maps (Bennett et al.\ \cite{benn:etal})
give
$R=0.35$ for the Galactic Center emission, and $R=0.17$ for the 
inner Galactic disk.
Following the derivation in the introduction,
the  excitation temperature of the \Jtwoone\ transition,
and hence the minimum
kinetic temperature of the \CI\ emitting gas in M\,82,
is $T_{\rm ex}\!=\!50 \mbox{ K}$.
The corresponding value from COBE for the Galactic Center is
$T_{\rm ex}\!=\!22 \mbox{ K}$, for the inner Galactic disk it is
$T_{\rm ex}\!=\!15 \mbox{ K}$.

Fig.\,2 shows the result of 
an escape
probability radiative transfer calculation (Stutzki \& Winnewisser
\cite{stu:gwi}) for spherical clumps, 
using the \Htwo\ collisional rate coefficients by
Schr\"oder et al.\ (\cite{schroe:etal}). 
It confirms the minimum temperature of 50\,K. 
At this temperature the line ratio 
implies a gas density of at least $10^4$ \cmcube.
At higher temperatures a lower density is possible 
($2\,10^3$ \cmcube\ at $T_{\rm kin}$=120\,K).
Densities higher than the quoted minimum require
the line to be optically thick and thermalized, and hence can only be
reached at very high CI-column densities (on the horizontal branch of
the according contour lines in Fig.\,2).
The minimum kinetic temperature and density derived are in good
agreement with the estimates for the excitation conditions of the bulk
CO emission from M\,82 (Turner et al.\ \cite{turner:etal}, Harris et al.\
\cite{harris:etal}, Wild et al.\ \cite{wild:etal}, 
G\"usten et al. \cite{guesten:etal}).
The \CI-line ratio constrains the
pressure of the interstellar medium in M\,82 to be $5\,10^5$ K
\cmcube\ or higher.

\begin{figure}
\plotfiddle{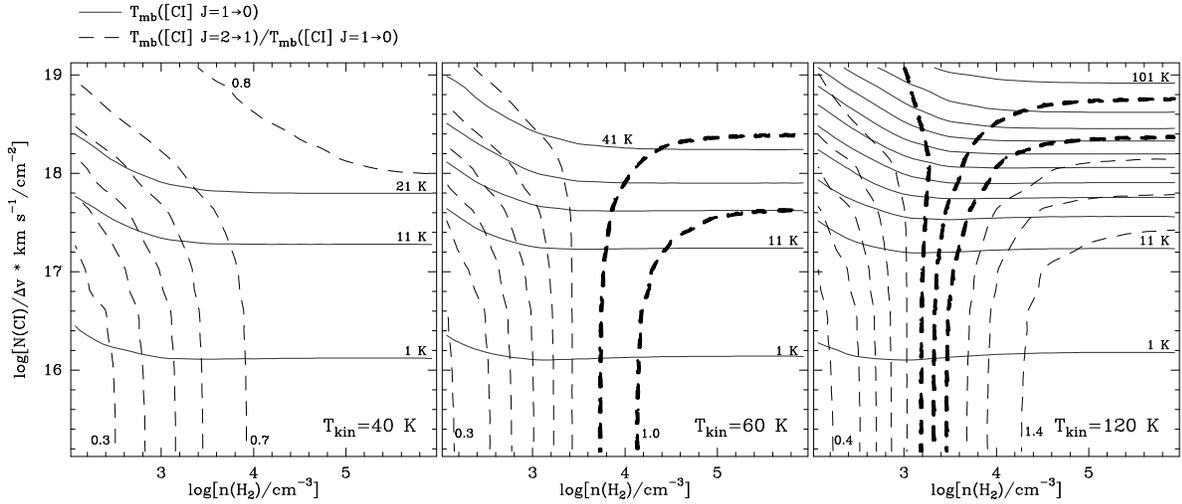}{6.5 cm}{270.}{63.}{63.}{-250}{230}
\caption{
Escape probability radiative transfer analysis of the
\CI\ emission:  \Jonezero\ 
line brightness temperature
(solid contours, increasing bottom to top) 
and the brightness ratio between the \Jtwoone\ and 
\Jonezero\ line
(dashed contours, increasing left to right) 
are plotted in a density/column density plane at 3
different kinetic temperatures. 
For
comparison with the
observed line ratio the ratio contours at 0.9, 1.0 and 1.1
are drawn as heavy dashed lines.
In the optically thin regime, a 1 K
bright \threePonezero\ line corresponds
to a beam averaged CI column density of $1.3\,10^{16}$ \cmsqu\ 
per 1 \kms velocity
interval. 
\label{fig2}
}
\end{figure}

\placefigure{fig2}

The absolute line
intensities correspond to a beam averaged CI column
density of
$2.1\,10^{18}$ \cmsqu. 
This value depends only weakly on the
temperature and density of the gas (see Fig.\,2). This is plausible, as the
partition function of the three level system CI does not change
significantly above a $T_{\rm ex}$ of about 20\,K, as was
noted already by Schilke et al.\ \cite{schilke:etal}.
Comparison with the total CO
column density of $4\,10^{18}$ \cmsqu\
derived by Wild et al.\ (\cite{wild:etal}) both from the \CeighteenO\
low-J intensities and an excitation model for all observed CO lines,
then gives a high CI/CO abundance ratio of 0.5, 
confirming the high value found by
Schilke et al.\ (\cite{schilke:etal}) and White et al.\ (\cite{white:etal}).

The reason for the high CI/CO ratio in M\,82 in comparison to
typical Galactic sources with a similar UV field intensity and comparable
densities, is unclear. It might primarily be a geometrical effect, where
a much smaller fraction of the presumably highly 
clumped  material is located at sufficient
column density to be well shielded from the UV radiation than
is the case for typical Galactic sources; the cause for this would, of
course, be linked to the rather different 
physical environment created by the starburst
activity in M\,82.
Schilke et al.\ \cite{schilke:etal}
argue that at the enhanced cosmic ray
flux in M\,82 (Suchkov et al.\ \cite{suchkov:etal}),
resulting from the high SN rate (Kronberg et al.\ \cite{kronb:etal})
following the
starburst, the high ionisation solution found by Le\,Bourlot et al.\ 
(\cite{lebour:etal}) 
for low density material in dark interstellar clouds
with its correspondingly large CI/CO ratio 
will be a stable solution of the chemical network
even at the high densities of the interstellar medium in M\,82.
The high
cosmic ray flux also
explains the high temperature of the ambient molecular cloud material.
St\"orzer et al.\ (\cite{stoe:etal}) 
point out that the CII layer in a PDR rapidly
recombines, but only slowly forms CO,
leading to a layer with enhanced CI abundance for
about $10^5$ years
when the illuminating UV
radiation is switched off.
Under repeated exposure to UV, as will naturally occur in a turbulent,
clumpy medium where the clumps mutually shadow each other, this will
on time average
lead to a significantly enhanced CI emission, in particular if the
medium stays warm, as will be the case in clouds like the Galactic
Center clouds or the interstellar medium in M\,82.
Which of these proposed mechanisms (and possibly others) provide the
right explanation for the enhanced CI abundance in M\,82 and other starburst
galaxies has to await
further
investigation. 

The
measured line ratio allows the determination of 
an upper limit to the CI column density as long as the
\CI\ emitting gas has temperatures below about 100\,K. 
Higher temperatures, 
however, are rather unlikely as there is no
known mechanism that would heat the bulk of the gas to such high
temperatures and yet keep the dust cool enough to be consistent with
the observed dust continuum emission. At temperatures
of about 50\,K the observed 
line ratio implies
an upper limit to the CI column density 
of about $4\,10^{19}$ \cmsqu\ (with the
equivalent width of the line of 100 \kms), 
20 times larger than the beam averaged column density. The
filling factor of individual clumps hence must be about 5\% or larger.
At
these high column densities both \CI\ lines 
become close to optically thick.

We now argue, that the \CI\ emitting gas in M\,82 originates most likely from
a large number of small clumps, where the individual clumps reach column 
densities close to the upper limit discussed above. With
{\it i)} a CI/CO abundance
ratio of 0.5, {\it ii)} a
total gas phase abundance of C/H of $10^{-4}$, {\it iii)} assuming
the gas phase carbon abundance to be dominated by CO and CI, and
{\it iv)} an \Htwo
volume density of about $10^4$ \cmsqu\ estimated from the excitation
analysis, a CI column density of $2.1\,10^{18}$ \cmsqu\ corresponds
to a geometric depth of the cloud of only 0.5 pc, or 0.03" at the distance 
of M\,82 of 3.25 Mpc (Tamman \& Sandage \cite{tam:sand}).
Even at the upper range of
column densities derived above, i.e. $1.5\,10^{19}$ \cmsqu\ for a giant
molecular cloud complex with a 30 \kms wide line, the linear
scale reaches only 3.6 pc, or about 0.22". This angular scale is consistent 
with the beam filling factor estimated above, assuming the emission to 
arise from on the order of 100 such complexes 
within the beam. In fact, structure is
visible down to at least 2" on interferometric maps both in CO \Jtwoone\ 
(Shen \& Lo \cite{shen:lo}) and
in the high density tracing molecule 
HCN (Brouillet \& Schilke \cite{broui:schi}).
Higher densities would imply smaller clump
sizes and hence a larger number of clumps in the beam. 
At higher temperatures  
the upper limit to the column density becomes higher (by an order of 
magnitude at 120\,K), and accordingly 
larger clump sizes are possible. The most likely scenario explaining the 
observed \CI\ line emission is thus an ensemble of cloud complexes with
linear sizes of a few parsecs and densities of several 
$10^4$ \cmcube, their temperatures ranging
between about 50\,K and up to about 100\,K.

In summary, our simultaneous observations of both
\CI\ fine structure lines at 809 and 492 GHz constrain the
physical conditions in the \CI-emitting gas to densities of $10^4$ \cmcube
or slightly higher and
temperatures of 50\,K, possibly up to 100\,K for a fraction of the gas. 
The analysis 
confirms the high CI/CO  abundance ratio of 0.5 in
the starburst galaxy M\,82. The emission arises in many cloud
complexes of a few parsecs extent,
which individually reach optical depths close to unity in the \CI\ lines.

\acknowledgments
This research was supported by the
{\it Verbundforschung Astronomie} through grant 05\,2KU134(6) 
and by
the {\it Deutsche Forschungsgemeinschaft} through grant SFB 301.
Special thanks go
to members of the SMT technical staff for their help during setup of the
instrument and interfacing it to the telescope.

\clearpage

\end{document}